\documentclass[twocolumn,prb,amsmath,amssymb,showpacs]{revtex4}

\usepackage{bm}

\ifx\pdftexversion\undefined
  \usepackage[dvips]{graphicx}
\else
  \usepackage[pdftex]{graphicx}
\fi

\def \be{\begin{equation}}
\def \ee{\end{equation}}
\def \bmlett{\begin{mathletters}}
\def \emlett{\end{mathletters}}

\def \pd{\phantom{\dagger}}

\def \ra{\rightarrow}

\def \sgn{{\rm sgn}}

\def \hrho{\hat{\rho}}
\def \hx{\hat{x}}
\def \hp{\hat{p}}

\def \tV{\tilde{V}}

\begin{document}



\title{Shot Noise of a Tunnel Junction Displacement Detector}

\author{A. A. Clerk and S. M. Girvin}
\affiliation{ Departments of Applied Physics and Physics, Yale
University, New Haven CT, 06511, USA\\
May 28, 2004}

\begin{abstract}
We study quantum-mechanically the frequency-dependent current
noise of a tunnel-junction coupled to a nanomechanical oscillator.
The cases of both DC and AC voltage bias are considered, as are
the effects of intrinsic oscillator damping.  The dynamics of the
oscillator can lead to large signatures in the shot noise, even if
the oscillator-tunnel junction coupling is too weak to yield an
appreciable signature in the average current. Moreover, the
modification of the shot noise by the oscillator cannot be fully
explained by a simple classical picture of a fluctuating
conductance.
\end{abstract}

 \maketitle

Spurred primarily by experiments in solid-state qubit systems,
there has recently been considerable interest in understanding the
noise properties of mesoscopic systems used as detectors
\cite{QPCPapers, MakhlinRMP, KorotkovSN, Pilgram, Me, Averin}.
Many new results have emerged, including an understanding of the
connection between noise, back-action dephasing and information
\cite{Pilgram, Me, Averin}, and of the influence of coherent qubit
oscillations on the output noise of a detector \cite{KorotkovSN}.
Not surprisingly, similar concerns arise in the study of
nanomechanical oscillators. Recent experiments using
single-electron transistors (SETs) have demonstrated displacement
detection of such oscillators with a precision close to the
maximum allowed by quantum mechanics \cite{Cleland,Schwab}. Given
the interest in these systems, it is important to gain a better
understanding of how a mesoscopic detector influences the
behaviour of an oscillator, and vice-versa. Several works have
addressed various aspects of this problem.  In particular, it has
been shown that an out-of-equilibrium detector can serve as an
effective environment for the oscillator, providing both a damping
coefficient and an effective temperature
\cite{MozTJ,Moz,Blencowe}.

In the present work, we turn our attention to the finite-frequency
output noise of a mesoscopic displacement detector, where one
expects to see signatures of the time-dependent fluctuations of
the oscillator. A completely classical study of the current noise
of a DC-biased SET displacement detector was presented recently in
Refs. \onlinecite{ArmourSI} and \onlinecite{BlanterSHO}.  In
contrast, we consider a generic tunnel-junction or quantum
point-contact (QPC) detector, in which the tunnelling strength
depends on the position of the oscillator, and calculate {\it
quantum mechanically} the finite frequency current noise.  Such a
system could be realized by using an STM setup where one electrode
is free to vibrate \cite{Yurke,Konrad}.  We treat both DC and AC
voltage bias; the latter is of particular interest, as in
experiment, it is common to imbed the detector in a resonant tank
circuit for impedance-matching purposes, and then probe its AC
response.  We find that even for a detector-oscillator coupling so
weak that there is little signature of the oscillator in the
average current, there can nonetheless be a strong signature in
the finite-frequency current noise. We moreover find that the
oscillator contribution to the noise cannot be simply explained by
a classical model of a detector conductance which fluctuates with
the oscillator position-- there are additional quantum corrections
which suppress the contribution of zero point fluctuations.  We
show that these quantum corrections result from correlations
between the detector's random back-action force and intrinsic
noise. Finally, in the AC-biased case, we find that the oscillator
experiences a {\it time-dependent} temperature, which has a direct
influence on the detector's current noise.

{\it Model}- Considering the simplest case where the tunnel-matrix
element depends linearly on the oscillator displacement $\hx$, the
tunnel junction detector is described by:
\begin{eqnarray}
    H_{det} & = & \frac{\tau_0 + e^{i \eta} \tau' \hx}{2 \pi \Lambda} \sum_{k,k'}
        \left(
            Y^{\dag} c^{\dag}_{R,k} c^{\pd}_{L,k'} + {\textrm h.c.}
        \right) - e V(t) \hat{m}
        \nonumber \\
        & \equiv & H_{det,0} - \hx \cdot \hat{F} \nonumber
\end{eqnarray}
Here, $c_{L,k}$ ($c_{R,k}$) destroys an electron state in the left
(right) electrode, $\Lambda$ is the conduction-electron density of
states, $\hat{m}$ denotes the number of tunnelled electrons, and
the operator $Y^\dag$ augments $\hat{m}$ by one.  $\eta$
parameterizes the sensitivity of the transmission phase to $\hx$,
and will in general be non-zero \cite{etaNote}. We consider both
the cases of a pure DC voltage, $V(t) = \tilde{V}$ and a pure AC
voltage, $V(t) = \tilde{V} \cos \nu t$. Note that the tunnelling
Hamiltonian itself acts as a random back-action force $\hat{F}$ on
the oscillator; this corresponds to random momentum shifts
imparted to the oscillator by tunnelling electrons \cite{Yurke}.
We will describe our system by a reduced density matrix
$\rho(m;x,x';t) \equiv \langle x | \hrho(m;t) | x' \rangle$ which
tracks the state of the oscillator and $m$, the number of
electrons which have tunnelled through the junction. As there is
no superconductivity, $\hrho$ is diagonal in $m$. In general, the
evolution of $\hrho$ will be given by a Dyson-type equation:
\begin{eqnarray}
    \frac{d}{dt} \hrho(m,t)  & = &
        -\frac{i}{\hbar}
        \left[ H_0, \hrho(m;t) \right]  +
            \int_{t_0}^{t} dt' \sum_{m'}
        \label{FullMaster2}
             \\ &&
        \check{\Sigma}(m,m';t-t') \circ
        \left[ U_0^{\pd}(t-t')
        \hrho(m';t') U_0^{\dag}(t-t')
        \right]  \nonumber
\end{eqnarray}
Here, $U_0$ is the evolution operator corresponding to the
unperturbed (zero-tunnelling) Hamiltonian, and we have written the
self-energy $\Sigma$ as a super-operator (i.e.~ an operator acting
on the space of density matrices).

We will consider the simplest case of weak tunnelling, and keep
only self-energy terms which are lowest order in the tunnelling.
$\Sigma$ is only non-vanishing if $m'=m$ or $m' = m \pm 1$;  these
two types of contributions correspond to ``scattering out" and
``scattering in" terms in a kinetic equation, and are given by the
diagrams shown in Fig. 1.  These diagrams correspond to standard
tunnelling bubbles \cite{Schoeller}, the only difference being
that the tunnelling vertices can contain an $\hx$ operator. If
$\hx$ appears at the $t'$ end of a graph for $\Sigma(t,t')$, $\hx$
will evolve during the duration of the tunnelling event.  As a
result, the self energy $\check{\Sigma}$ has terms involving
$\hp$, and the final form of $\Sigma$ we obtain {\it does not}
correspond to the oscillator-free case with $\hx$ dependent rates.
We also include perturbatively the effects of a high-temperature
Ohmic heat bath ($k_B T_{bath} \gg \hbar \Omega$, with $\Omega$
being the oscillator frequency) on the oscillator using a
Caldeira-Leggett description \cite{CL} and the lowest-order Born
diagrams in the self-energy (i.e.~ same diagrams as in Fig. 1,
with tunnelling bubbles replaced by environmental boson lines).

\begin{figure}
\center{\includegraphics[width=7.5 cm]{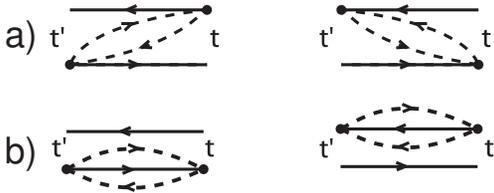}} \vspace{-0.5 cm}
\caption{\label{QLPlot1} Diagrams for the a) scattering in, and b)
scattering out terms in the self energy $\check{\Sigma}(t-t')$.
The solid lines represent the forward and backwards Keldysh
contours; the dashed lines are conduction electron propagators.
The solid black vertices correspond to $\tau_0 + \tau' \hx$.  Note
that an $\hx$ operator appearing in a vertex at time $t'$ will
evolve during the tunnelling event. } \vspace{-0.5 cm}
\end{figure}

Finally, we specialize to the case where the voltage $\tilde{V}$
is much larger than $\hbar \Omega / e$.  For weak tunnelling, $e
\tV \gg \hbar \Omega$, and small AC frequency $\nu$, it is then
possible to make a Markov approximation in
Eq.~(\ref{FullMaster2}): $
     U_0^{\pd}(t-t')
        \hrho(m';t') U_0^{\dag}(t-t')
    \ra
     \hrho(m';t)$.
We are assuming that over the short timescales relevant to
tunnelling, one can describe the dynamics of the density matrix by
its zero-tunnelling evolution. Fourier-transforming in the $m$
index, $
    \hrho(k;t) = \sum_{m=-\infty}^{\infty} e^{i k m}
        \hrho(m;t)$,
Eq.~(\ref{FullMaster2}) becomes:
\begin{widetext}
\begin{eqnarray}
    \frac{d}{dt} \hrho(k;t) & = &
                -\frac{i}{\hbar} \left[ H_0 - \bar{F}(t,\eta) \hx , \hrho \right]
        - i \left(\frac{\gamma_0 + \gamma}{\hbar} \right)
            [ \hx, \{ \hp, \hrho \} ]
        -  \left(\frac{D_0 + D(t)}{\hbar^2}\right) [ \hx, [ \hx, \hrho ] ]
        \label{MassiveEqnAC}
      + \sum_{\sigma = +,-}
            \left(
                \frac{e^{i \sigma  k}-1}{ (\tau')^2}
            \right)
      \times  \\  &&
           \Bigg(
        \frac{2 D_{\sigma}(t)}{\hbar^2 }
                (\tau_0 + e^{i \sigma \eta} \tau'  \hx) \hrho
                    (\tau_0 + e^{-i \sigma \eta} \tau'  \hx)
                \nonumber
        + i \frac{\gamma_{\sigma}(t)}{\hbar} \left(
            \tau_0 \tau' \left( e^{i \sigma \eta} \hp \hrho -
             e^{-i \sigma \eta} \hrho \hp \right) +
            (\tau')^2 \left( \hp \hrho \hx - \hx \hrho \hp \right)
            \right) \Bigg) \nonumber
\end{eqnarray}
\end{widetext}
where $\gamma_0$ is the intrinsic damping coefficient associated
with the equilibrium bath, $D_0 = 2 M \gamma k_B T_{bath}$ is the
corresponding diffusion constant, and $\sigma = +(-)$ labels
contributions from forward (backwards) tunnelling. The
detector-dependent diffusion constant $D(t) = \sum_{\sigma}
D_{\sigma}(t)$ and damping coefficient $\gamma(t) = \sum_{\sigma}
\gamma_{\sigma}(t)$ are given by:
\begin{eqnarray}
    \gamma_\sigma(t) & = &
    \frac{\hbar }{2 M \Omega}
    \left(\frac{\tau'}{\tau_0}\right)^2
        \left(
            \frac{\Gamma_{\sigma}(t,\hbar \Omega)
                -\Gamma_{\sigma}(t,-\hbar \Omega)}{2}
            \right)
            \label{GenGamma}\\
    D_{\sigma}(t) & = &
        \frac{\hbar^2}{4}
        \left( \frac{\tau'}{\tau_0} \right)^2
       \left(
            \Gamma_{\sigma}(t,\hbar \Omega)
                +\Gamma_{\sigma}(t,-\hbar \Omega)
            \right)
            \label{GenDiff}
\end{eqnarray}
while $\bar{F}(t,\eta) = \sin \eta
\left(\frac{\tau'}{\tau_0}\right) \sum_{\sigma} 2 \sigma
D_{\sigma}(t) / \hbar$ is the average back-action force exerted on
the oscillator. $\Gamma_{\pm}(t,E)$ are the $\tau'=0$ finite
temperature forward and backwards inelastic tunnelling rates
involving an absorbed energy $E$; these rates are time-independent
in the case of a DC voltage. Note that we have neglected
self-energy terms which renormalize the oscillator Hamiltonian;
these are unimportant in the weak-tunnelling limit we consider.

Eq.~(\ref{MassiveEqnAC}) yields a compact description of the
coupled detector-oscillator system; it is a generalization of an
equation first derived (via an alternate approach) by Mozyrsky
{\it et al.} \cite{MozTJ} to an {\it arbitrary} detector in the
tunnelling regime, including the possibility of an $x$-dependent
tunnelling phase\cite{etaNote}, a nonlinear junction I-V, a
time-dependent bias voltage, and intrinsic oscillator damping.
Taking $k=0$ yields the equation for the reduced-density matrix of
the oscillator, and (c.f.~ Ref. \onlinecite{MozTJ}) has the
Caldeira-Leggett form for a forced, damped oscillator in the
high-temperature regime \cite{CL}. In what follows, we focus for
simplicity on the case of $T=0$ in the tunnel junction, and on
$\eta = 0$, which ensures $\bar{F}=0$ \cite{etaNote}; a non-zero
$\bar{F}$ does not significantly change our results.

{\it Shot Noise- }Eq.~(\ref{MassiveEqnAC}) can in principle be
used to calculate the full counting statistics of tunnelled charge
as a function of time.  By focusing solely on the time-dependence
of the reduced second moment $\langle \langle m^2(t) \rangle
\rangle$ (i.e.~ variance), it is possible to calculate the
symmetrized frequency-dependent current noise using the MacDonald
formula \cite{MacDonald}.  In the case of an AC bias voltage, the
noise will be a function of two times. We focus on the part of the
noise that is independent of the average time co-ordinate, a
quantity which is directly accessible in experiment.  It is given
by a modified version of the MacDonald formula:
\begin{equation}
    S_I(\omega) = 2 e^2 \omega
        \int_0^{\infty}dt
            \sin \omega t
        \int_0^{2 \pi} \frac{d \phi}{2 \pi}
    \cdot \partial_t \langle \langle m^2(t,\phi) \rangle \rangle
    \label{MacDonald}
\end{equation}
where $\phi$ is the initial phase of the AC voltage.

{\it DC Bias}- For a DC biased normal-metal junction at $T=0$, the
tunnelling rates are given by $h \Gamma_{\sigma}(t,E) = (\tau_0)^2
(\sigma e \tV + E) \Theta(\sigma e \tV + E)$. Eqs.
(\ref{GenGamma})-(\ref{GenDiff}) yield $\gamma = \hbar \tau'^2/ (4
\pi M)$ and $k_B T_{eff} = e V/2$ \cite{MozTJ}.  We find from Eqs.
(\ref{MassiveEqnAC}) and (\ref{MacDonald}) that the current noise
may be written as $S_I(\omega) = 2 e \langle I \rangle + \Delta
S_I$, where the first term corresponds to purely Poissonian
statistics, and the second term is a correction arising from
correlations between the motion of the oscillator and the number
of tunnelled electrons:
\begin{eqnarray}
    \Delta S_I (\omega) & = &
            \frac{4 e^3 \tV}{h} \omega
        \int_0^{\infty} dt \sin \omega t \hspace{3 pt}
         \label{DSI} \\
            &&
        \Big(
            \left(2 \tau_0 \tau' \right)
                \langle \langle \hx(t) \cdot m(t) \rangle \rangle
                + \left(\tau' \right)^2
                \langle \langle \hx^2(t) \cdot m(t) \rangle \rangle
        \Big) \nonumber
\end{eqnarray}
Physically, the covariances appearing above arise from the
$x$-dependence of the tunnelling probability-- if $m(t)$ is larger
than average, then it is likely that $x(t)$ and $x^2(t)$ are also
larger than average.  These covariances can be calculated directly
from Eq.~(\ref{MassiveEqnAC}), and obey simple classical equations
corresponding to a forced, damped harmonic oscillator. Consider
first the contribution from $\langle \langle x \cdot m \rangle
\rangle$ in Eq.~(\ref{DSI}), which is leading order in $\tau'$. In
calculating this covariance, one finds that the tunnel junction
provides an effective driving force; we find a contribution:
\begin{eqnarray}
    \Delta S_I(\omega) \Big|_{1} = \frac{e^3 \tV}{h}
        \left( 2 \tau_0 \tau' \right)^2
            \left( \frac{e \tV}{h}  -
                \frac{\Omega}{4 \pi} \frac{(\Delta x_0)^2}{\langle x^2 \rangle}
            \right)
                S_x(\omega)
        \label{DSI1}
\end{eqnarray}
where $S_x(\omega) = 8 \gamma \Omega^2 \langle x^2\rangle / (
(\omega^2 - \Omega^2)^2 + 4 \gamma^2 \omega^2)^{-1}  $ is the
spectral density of oscillator $x$ fluctuations obtained from
Eq.~(\ref{MassiveEqnAC}), and $(\Delta x_0)^2 = \hbar/(2 M
\Omega)$ is the zero-point uncertainty in the oscillator position.
The first term in Eq.~(\ref{DSI1}) is {\it exactly} the answer
expected (to lowest order in $\tau'$) from a simple picture of a
classically fluctuating junction conductance (i.e.~ $\Delta
S_I(\omega) = \tV^2 S_G(\omega)$, where $S_G(\omega)$ is the
spectral density of conductance fluctuations, and is in turn
determined by $S_x(\omega)$). Equivalently, if we think of our
junction as an $x$-to-$I$ amplifier having a gain $\lambda = 2 e
\tV \tau_0 \tau'/h$, this first term corresponds to simply
amplifying up the fluctuations of the oscillator: $\Delta S_I =
\lambda^2 S_x$. Eq.~(\ref{DSI1}) yields a peak in $S_I(\omega)$ at
$\omega = \Omega$; keeping only the leading term in $\tV$, the
ratio of the peak-height to the background Poissonian noise (i.e.~
the S/N ratio) is:
\begin{equation}
    \frac{\Delta S_I(\omega=\Omega)}{2 e \langle I \rangle}  =
        4 \tau_0^2 \left(\frac{e \tV}{h \gamma_{tot}}\right)
            \frac{\alpha^2}
            {1 +  \alpha^2 }
            \leq
            4 \left(\frac{ \tau_0}{\tau'}\right)^2
             \frac{2 M e \tV}{\hbar^2}
             \label{SNEqn}
\end{equation}
where $\alpha^2 = \tau'^2 \langle x^2 \rangle / \tau_0^2$,
$\gamma_{tot} = \gamma_0 + \gamma$. Note that if $\alpha$ is
small, there will be no sizeable signature of the oscillator in
the average current (i.e.~ $\delta \langle I \rangle / \langle I
\rangle_0 \simeq \alpha^2 $), but there may nonetheless be a large
peak in the noise if $e \tV / (h \gamma_{tot})$ is large. The
upper bound in Eq.~(\ref{SNEqn}) corresponds to the optimal
scenario, where there is no intrinsic (detector-independent)
damping, and $\alpha \gg 1$. The maximum $S/N$ is determined by $e
V$ and the sensitivity $\tau'/ \tau_0$, and can be arbitrarily
large.  Due to the dependence on $\gamma$, we find the surprising
result that the maximum $S/N$ is {\it inversely} proportional to
the detector sensitivity $\tau' / \tau_0$.  Note the marked
difference from experiments attempting to detect coherent qubit
oscillations in the detector current noise \cite{KorotkovSN},
where back-action effects limit the $S/N$ to a maximum of $4$.

We turn now to the second term in Eq.~(\ref{DSI1}), which is a
lower-order in $\tV$ quantum correction to the classical result.
It would appear to cause $\Delta S_I |_1$ to vanish in the limit
$e \tV \ra \hbar \Omega / 2$, $\langle x^2 \rangle \ra (\Delta
x_0)^2$, i.e.~ it suppresses a zero-point contribution to $\Delta
S_I |_1$. (Of course, we cannot rigorously take this limit, as
Eq.~(\ref{MassiveEqnAC}) is strictly only valid for $e \tV \gg
\hbar \Omega$). A similar result was found for the average current
$\langle I \rangle$ in Ref. \onlinecite{MozTJ}, where a similar
offset term could be traced to the inherent asymmetry between
events in which energy is absorbed from the oscillator, versus
those in which it is emitted to the oscillator. In the present
case, the quantum correction to the noise in Eq.~(\ref{DSI1}) can
be given a classical interpretation-- it arises from correlations
between the intrinsic shot noise of the detector, and the
back-action force $\hat{F}$ acting on the oscillator. If there are
such correlations, we would expect classically the current noise
to have the form:
\begin{equation}
    \Delta S_I(\omega) = \lambda^2 S_x(\omega) +
        2 \lambda \textrm{Re } \left[ g(-\omega) S_{I F}(\omega) \right]
\end{equation}
where $S_{I F}(\omega)$ is the symmetrized cross-correlator
between the junction current and back-action force noise, and
$g(\omega)$ is the oscillator response function.  Note that the
second term above is $\propto \tV$, while the first is $\propto
\tV^2$.  It is well known that inversion symmetry forces
$\textrm{Re } S_{I F}$ to vanish; this is what allows a QPC
detector to reach the quantum limit for measuring a qubit
\cite{Me,Averin}. However, at finite $\omega$, $\textrm{Im } S_{I
F}$ is non-zero. Consequently, the second term above is non-zero;
a direct perturbative calculation (assuming a thermal state for
the oscillator) shows that this term corresponds to the second
term in Eq.~(\ref{DSI1}). Thus, we see that {\it quantum
corrections to the noise, which suppress zero-point contributions,
can be associated with classical out-of-phase correlations between
the random back-action force and the intrinsic detector output
noise}.

Finally, we return to Eq.~(\ref{DSI}) and examine the contribution
from $\langle \langle x^2 \cdot m \rangle \rangle$, a term which
is higher-order in $\tau'$.  One finds:
\begin{eqnarray}
    \left[ \Delta S_I(\omega) \right]_2 & = &
         \frac{e^3 \tV}{h} (\tau')^4
            \left( \frac{e \tV}{h}  -
                \frac{\Omega}{2 \pi} \frac{(\Delta x_0)^2}{\langle x^2 \rangle}
            \right)
            \nonumber \\
         && \times
         \int \frac{d \omega'}{2 \pi}
            S_x(\omega') S_x(\omega - \omega')
            \label{DeltaSI2Eqn}
\end{eqnarray}
Again, the first term above agrees with the expectation for a
classically fluctuating junction conductance; it yields peaks in
$S_I$ at $\omega = 0$ and $\omega = 2 \Omega$.  The second term is
a quantum correction, completely analogous to that found for
$\Delta S_I |_1$.

\begin{figure}[t]
\center{\includegraphics[width=7 cm]{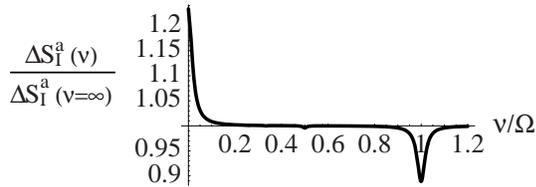}} \vspace{-0.4
cm}\caption{\label{SIConst} Oscillator contribution to $S_I^a$,
the frequency-independent part of the shot noise  (i.e.~ second
term in Eq.~(\ref{SIFreqIndep})), versus the AC voltage frequency
$\nu$, for $\Omega / \gamma = 50, \gamma_0=0$, and $e \tilde{V}
\gg \hbar \nu, \hbar \Omega$. The maximum suppression of this term
at $\nu = \Omega$ (over its $\nu \ra \infty$ limit) is by 8/9.}
\end{figure}
\begin{figure}[t]
\center{\includegraphics[width=7.5 cm]{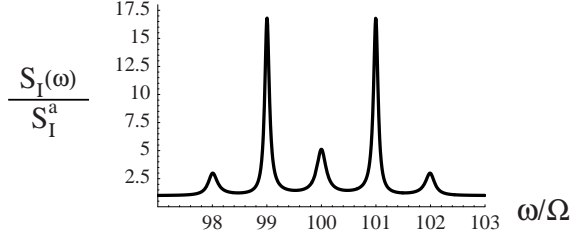}} \vspace{-0.5
cm} \caption{\label{SICorrPlot} Full shot noise $S_I(\omega)$ for
an AC bias voltage of frequency $\nu = 100 \Omega$, including the
effects of correlations between $x(t)$ and $m(t)$ (c.f Eq.
(\ref{ACCorrNoise})).  We have chosen $e \tV = 100 \hbar \Omega$,
$\alpha^2 = 1$, $(\gamma_0 + \gamma) = \Omega / 20$, and $\tau_0 =
0.1$.  The y-axis is scaled by the value of the
$\omega$-independent part of the noise.}
\end{figure}
{\it AC Bias}- We now consider an AC bias voltage $V(t) = \tV
\cos(\nu t)$, where $e \tV \gg \hbar \nu, \hbar \Omega$. In the
limit of small $\nu$, it is possible to derive a simple expression
for the time-dependent tunnelling rates \cite{Tucker}. Defining $h
\tilde{\Gamma}(E) = (\tau_0)^2 E \cdot \Theta(E)$, we have $
\Gamma_{\sigma}(t,E)  =  \sum_{n=0}^{\infty}
        \left(1 - \delta_{n,0}/2 \right) \sigma^n
        \Gamma^{(n)}_{\sigma}(E) \cos n \nu t$, with:
\begin{eqnarray}
    \Gamma_\sigma^{(n)}  & = &
        \sum_{\pm}
        \int_0^{\pi} \frac{d \theta}{\pi} \cos(n \theta)
            \tilde{\Gamma}
            \left(
                e \tilde{V} \cos \theta +
                    E \pm \frac{n \hbar \nu}{2}
                \right)
        \label{LambdaKern}
\end{eqnarray}
Using Eqs. (\ref{GenGamma})-(\ref{GenDiff}), we find that the
damping coefficient $\gamma$ of the oscillator is time-independent
and identical to that in the DC case, whereas the diffusion
constant is time-dependent and contains higher harmonics of the AC
frequency $\nu$. Writing $D(t) = 2 M \gamma k_B T_{eff}(t)$, we
have to a good approximation:
\begin{eqnarray}
     k_B T_{eff}(t)
        & = & \frac{ e \tV}{\pi}
            \left(\sum_{n=0}^{e \tV / \hbar \nu}
            \left(\frac{2 (-1)^n}{1-(2 n)^2}\right)
          \cos(2 n \nu t) - 1\right)
         \nonumber \\
\end{eqnarray}
The small but finite photon frequency $\nu$ prevents higher
harmonics from contributing to $T_{eff}$; without it, we would
have simply $k_B T_{eff}(t) = \tV | \cos \nu t | / 2$, which tends
to zero twice each period.  With the finite cut-off included, the
minima of $k_B T_{eff}(t)$ are $\simeq \hbar \nu$.  The
time-dependence of $T_{eff}(t)$ implies that the position variance
$\langle x^2(t) \rangle$ of the oscillator will be time-dependent
and in phase with the AC voltage; as we show, this has a direct
influence on the noise and the average current. For the latter
quantity, we find:
\begin{eqnarray}
    \langle I(t) \rangle
    & = &
    \frac{e^2 \tV}{h} \cos(\nu t) \left(
            \tau_0^2 + (\tau')^2 \langle x^2(t) \rangle \right)
         - \Delta I(t)
         \nonumber
\end{eqnarray}
where the quantum correction is approximately $\Delta I(t) \simeq
e \gamma \cdot \sgn \left[\cos \nu t \right]$. Turning to the
noise, we may again decompose $S_I(\omega)$ into a
frequency-independent part and a term arising from correlations
between $x(t)$ and $m(t)$:
 $   S_I(\omega)  \equiv
        \frac{\Omega}{2 \pi}
        \int_0^{\frac{2 \pi}{ \Omega}}
        d \bar{t}
        S_I(\bar{t},\omega) =
        S_{I}^{a} + \Delta S_I(\omega) $.
For the frequency-independent contribution $S_{I}^a$, we find:
\begin{equation}
    S_{I}^a
         =
            \frac{4 e^2}{h} \Bigg[
            \frac{e \tV}{\pi}  \tau_0^2
            + \label{SIFreqIndep}
            (\tau')^2 \left(
        \overline{ k_B T_{eff}(t) \langle x^2(t) \rangle }
 - \frac{\Omega}{2 \pi} (\Delta x_0)^2 \right)
    \Bigg]
\end{equation}
where the bar indicates a time-average.  The first term is the
standard result for the shot noise of an AC-biased junction
\cite{PATNoise}. The second term indicates that the time-dependent
motion of $\langle x^2(t) \rangle$ (calculated from
Eq.~(\ref{MassiveEqnAC})) can make a frequency-independent
contribution to the noise.  For $\nu \gg \Omega$, $\langle x^2
\rangle$ responds only weakly to the time-dependence of
$T_{eff}(t)$, whereas for $\nu \sim \Omega$, the response becomes
appreciable and $180$ degrees out-of-phase with $V(t)$.  If in
addition $\gamma_0 \ll \gamma$, one finds a resulting {\it
suppression} of the oscillator's contribution to $S_I^a$; this is
shown in Fig. 2. Small resonances also occur when $\Omega$ is a
multiple of $\nu$. Note that the oscillator modification of
$S_I^a$ is not captured by the classical picture of a fluctuating
conductance.

Finally, the frequency-dependent contribution $\Delta S_I(\omega)$
to the noise, which arises from correlations between $x(t)$ and
$m(t)$, takes the simple form:
\begin{equation}
  \Delta S_I(\omega)   =  \frac{1}{4}\sum_{\pm}
    \Delta S_I(\nu \pm \omega) \Big|_{DC}
    \left[
        1    + O\left(\frac{\hbar \Omega}{ e \tV} \right)
    \right]
    \label{ACCorrNoise}
\end{equation}
where the omitted terms correspond to ``quantum corrections" of
the sort previously discussed.  Without these terms,
Eq.~(\ref{ACCorrNoise}) is precisely the answer expected for a
fluctuating classical conductance-- one needs to simply shift the
noise in the DC case up to the frequency $\nu$.  In contrast, the
quantum corrections to $\Delta S_I$ for AC bias are not simply
given by shifting the corresponding terms found for DC bias-- one
finds that the quantum corrections are larger in the AC case by a
factor of $4/\pi$. The effect of $\Delta S_I(\omega)$ on the full
noise is shown in Fig. 3.

In conclusion, we have presented a fully quantum mechanical
calculation of the frequency-dependent current noise of a tunnel
junction displacement detector, for both the cases of DC and AC
voltage bias. The oscillator can lead to large effects in the shot
noise, even if the coupling to the detector is weak; moreover,
these effects cannot be completely described using a classical
picture of a fluctuating junction conductance.  We thank Konrad
Lehnert and Florian Marquardt for useful discussions; this work
was supported by the NSF under grant NSF-ITR 0325580, and by the
W. M. Keck Foundation.


\begin{thebibliography}{99}
\bibitem{etaNote}
    For example, in a tunnelling
    setup where the oscillator modulates the width of a
    rectangular barrier by moving the position of the
    right electrode (c.f. Ref. \onlinecite{Yurke}), one finds
    $\tan^2 \eta = E_F / W$, where $W$ is the barrier height and $E_F$ the
    Fermi energy.
\end{thebibliography}
\end{document}